\documentclass[aps,prl,twocolumn,superscriptaddress,showpacs]{revtex4-1}
\makeatletter

\newcommand{\Rmnum}[1]{\expandafter\@slowromancap\romannumeral #1@}
\makeatother

\usepackage{graphicx}
\usepackage{latexsym}
\usepackage{amssymb}
\usepackage{amsfonts}
\usepackage{soul}
\usepackage{color}
\usepackage{amsmath}
\usepackage{natbib}
\usepackage[english]{babel}

\begin{document}

\title{Paramagnetic Meissner effect in ZrB$_{12}$ single crystal with non-monotonic vortex-vortex interactions}

\author{Jun-Yi Ge}\thanks{Junyi.Ge@kuleuven.be}
\affiliation{INPAC -- Institute for Nanoscale Physics and Chemistry, KU Leuven, Celestijnenlaan 200D, B--3001 Leuven, Belgium}
\author{Vladimir N. Gladilin}
\affiliation{INPAC -- Institute for Nanoscale Physics and Chemistry, KU Leuven, Celestijnenlaan 200D, B--3001 Leuven, Belgium}
\affiliation{TQC--Theory of Quantum and Complex Systems, Universiteit Antwerpen, Universiteitsplein 1, B--2610 Antwerpen, Belgium}
\author{Nikolay E. Sluchanko} 
\affiliation{Prokhorov General Physics Institute of RAS, 119991 Moscow, Russia}
\author{A. Lyashenko}
\affiliation{Frantsevich Institute for Problems of Materials Science, NASU, 03680 Kiev, Ukraine}
\author{Volodimir B. Filipov}
\affiliation{Frantsevich Institute for Problems of Materials Science, NASU, 03680 Kiev, Ukraine}
\author{Joseph O. Indekeu}
\affiliation{Institute for Theoretical Physics, KU Leuven, Celestijnenlaan 200D, B-3001 Leuven, Belgium}
\author{Victor V. Moshchalkov}
\affiliation{INPAC -- Institute for Nanoscale Physics and Chemistry, KU Leuven, Celestijnenlaan 200D, B--3001 Leuven, Belgium}

\date{\today}
\begin{abstract}

The magnetic response related to paramagnetic Meissner effect (PME) is studied in a high quality single crystal ZrB$_{12}$ with non-monotonic vortex-vortex interactions. We observe the expulsion and penetration of magnetic flux in the form of vortex clusters with increasing temperature.  A vortex phase diagram is constructed which shows that the PME can be explained by considering the interplay among the flux compression, the different temperature dependencies of the vortex-vortex and the vortex-pin interactions, and thermal fluctuations. Such a scenario is in good agreement with the results of the magnetic relaxation measurements.

\end{abstract}



\maketitle

\section{I. INTRODUCTION}
The perfect diamagnetism in a low enough magnetic field (Meissner effect) together with the zero resistivity are two most essential phenomena of superconductivity. However, it is found that a positive magnetic moment can appear below the critical temperature during field-cooling (FC), which is called paramagnetic Meissner effect (PME) or Wohlleben effect \cite{Li}. The PME was first reported in a polycrystalline high-temperature superconductor (HTS) Bi$_2$Sr$_2$CaCu$_2$O$_{8+\delta}$ by Svedlindh et al \cite{Svedlindh}. Then it was interpreted by the $d$-wave mechanism of HTS where the existence of  $\pi$-junctions characterized by negative Josephson couplings may lead to the positive magnetization \cite{Sigrist}. Later, the discovery of PME in single crystals of Nb \cite{Thompson,Kostic} and Al \cite{Geim}, which are all conventional superconductors, implies an alternative possible mechanism for PME. Koshelev and Larkin suggested that the positive magnetization is due to a Bean state with compressed magnteic flux \cite{Koshelev}. Based on the numerical solution of the Ginzburg-Landau equations, Moshchalkov et al. \cite{Moshchalkov} proposed that the PME arises from the flux compression with integral number of quantum flux $L\Phi_0$, where $\Phi_0$ is the flux quantum, trapped in the sample interior. Later on, by studying the PME in a mesoscopic superconductor Geim et.al. found that the quantized flux trapped at the third critical field is responsible for the PME \cite{Geim}, which supports the theoretical prediction in Ref.\cite{Moshchalkov}. The flux compression mechanism seems to be more universal to explain the PME observed both in HTS and conventional superconductors.

However, besides the possible interpretations mentioned above, the vortex state in the presence of the PME is still unclear. It is reported that not all the samples show PME even with similar nominal composition \cite{Okram,Felner}. Also the PME disappears after polishing the surface of the sample \cite{Kostic}, indicating that the surface configurations, like the defects and pinning centers, play an important role in the PME. Recently, it has been proposed that,  in multi-band superconductors with broad crossover from traditional type-I to type-II superconductivity (also termed as type-II/1 regime), giant PME may appear due to the non-monotonic vortex interactions and multibody effects \cite{Silva}, which may facilitate the trapping of magnetic flux. So far, experimental evidence of PME in type-II/1 superconductors is still lacking. 

In this article, as an effort to understand the vortex behavior in a superconductor with non-monotonic vortex interactions, we studied the PME in a high quality ZrB$_{12}$ single crystal ($2.48 \times 2.48 \times 0.72~mm^3$). The ZrB$_{12}$ crystal has a Ginzburg-Landau parameter of $\kappa \approx 0.8$, placing the sample right within the traditional type-I and type-II crossover regime \cite{Sluchanko,Ge-ZrB12}. The non-monotonic vortex interactions have also been confirmed by direct visualization of the very inhomogeneous vortex pattern at low temperatures\cite{Ge-ZrB12}. Clearly observed surface superconductivity of ZrB$_{12}$\cite{Leviev, Tsindlekht}  favors the formation of PME according to the flux compression mechanism. Moreover, the upper critical field is low ($\sim$ 600 Oe in Ref.\cite{Wang}). This allows us to study the vortex behavior near the phase boundary even at low temperatures. All the properties mentioned above make ZrB$_{12}$ a perfect platform to study the PME. 

\begin{figure*}[htb]
\centering
\includegraphics*[width=0.65\linewidth,angle=0]{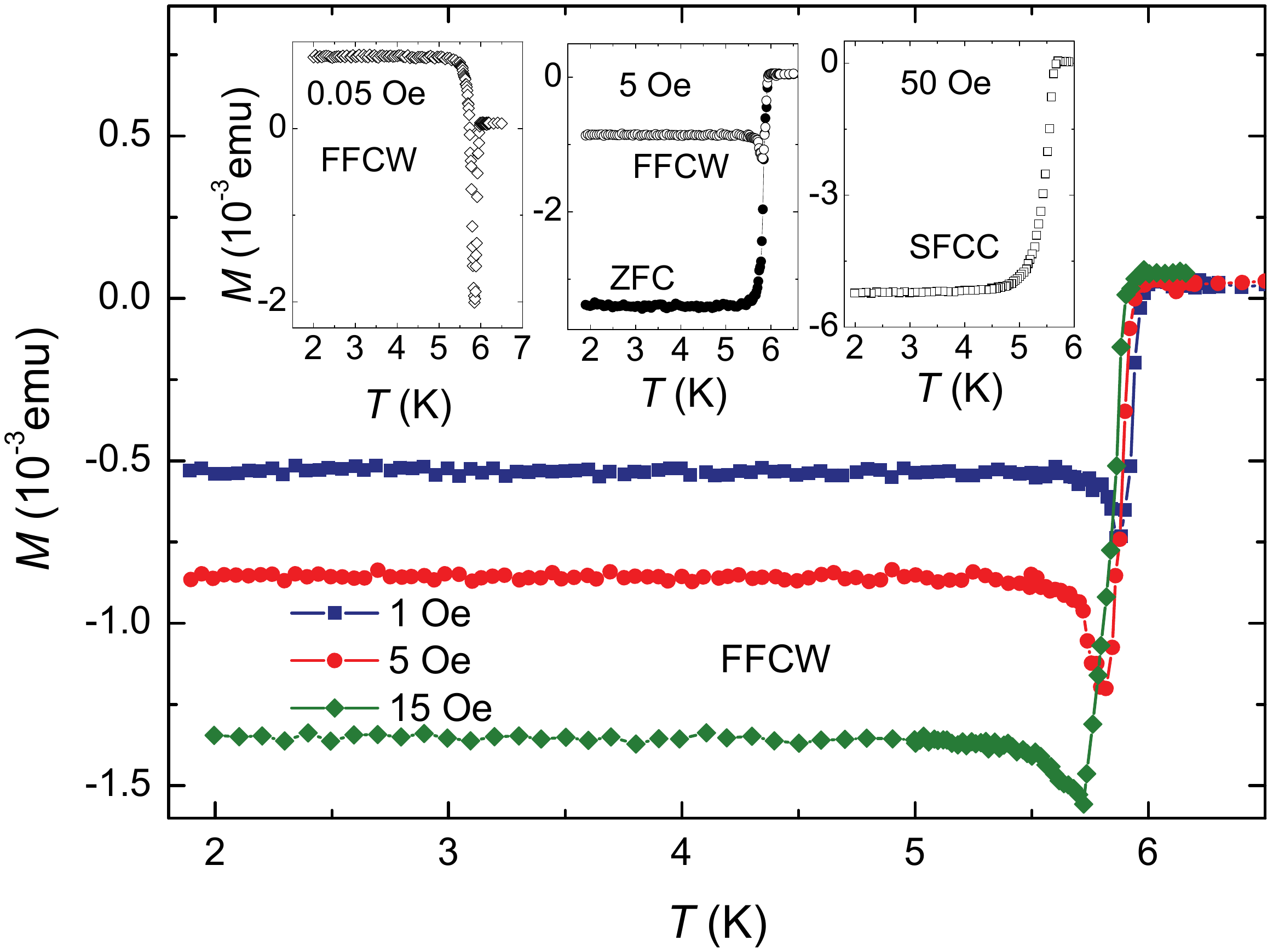}
\caption{(Color online) Fast-Field-Cool-Warming curves under magnetic fields of 1 Oe, 5 Oe and 15 Oe. The left inset shows an FFCW curve at 0.05 Oe, where positive magnetization signal is clearly observed. The middle inset shows the ZFC and FFCW curves at $H=5$ Oe. The right inset shows the SFCC curve at $H=50$ Oe.} 
\end{figure*}

We observed the expulsion of quantized flux trapped in the sample interior with increasing temperature after rapid field cooling and the penetration of flux in the form of  vortex clusters in the thermal fluctuation regime. The complex phase diagram is constructed and the magnetic relaxation effect is studied, both of which support a transition from vortex solid to vortex liquid through a regime of strong thermal fluctuations. As a result, the PME can be understood by considering the interplay of the following mechanisms: 1) flux compression; 2) temperature dependence of vortex-pin interactions and vortex-vortex interactions; 3) thermal fluctuations.

\section{II. EXPERIMENTAL}
The high quality ZrB$_{12}$ single crystal was grown by the inductive zone melting method, with critical temperature $T_\textrm{c}=5.95$ K and a transition width of 0.08 K under an external field of 1 Oe. Details of the sample preparation can be found in Ref.\cite{Daghero}. The temperature dependence of the magnetization was measured using the following preparations: 1) $Zero$-$Field$-$Cooling$ (ZFC), the sample was initially cooled in the absence of a magnetic field to 2 K, and subsequently the magnetization was measured with increasing temperature under a magnetic field of $H$. 2) $Fast (Slow)$-$Field$-$Cool$-$Warming$ (F(S)FCW), the sample was cooled with a large (small) cooling rate 5 K/min (0.03 K/min) to the required temperature under a magnetic field of $H$, then the magnetization was measured with increasing temperature. 3) $Slow$-$Field$-$Cool$-$Cooling$ (SFCC), the magnetization was measured with decreasing temperature at a rate of 0.03 K/min to the desired temperature under different magnetic fields. The magnetic moments and the ac susceptibility measurements were performed by using the Quantum Design PPMS with an ACMS option in both dc and ac modes. To avoid the thermal gradient effect that might leads to the discrepancy between FCC and FCW curves, the magnetic moment data were collected at a small sweeping rate (0.1 K/min). The vortex patterns were visualized using Low Temperature Scanning Hall Probe Microscopy (SHPM) from Nanomagnetics Instruments as introduced in Ref.\cite{Ge-NC,Ge-NC2}. The dc and ac fields were applied along the (110) direction, which is perpendicular to the sample surface.

\section{III. RESULTS AND DISCUSSIONS}

\begin{figure*}[htb]
\centering
\includegraphics*[width=0.75\linewidth]{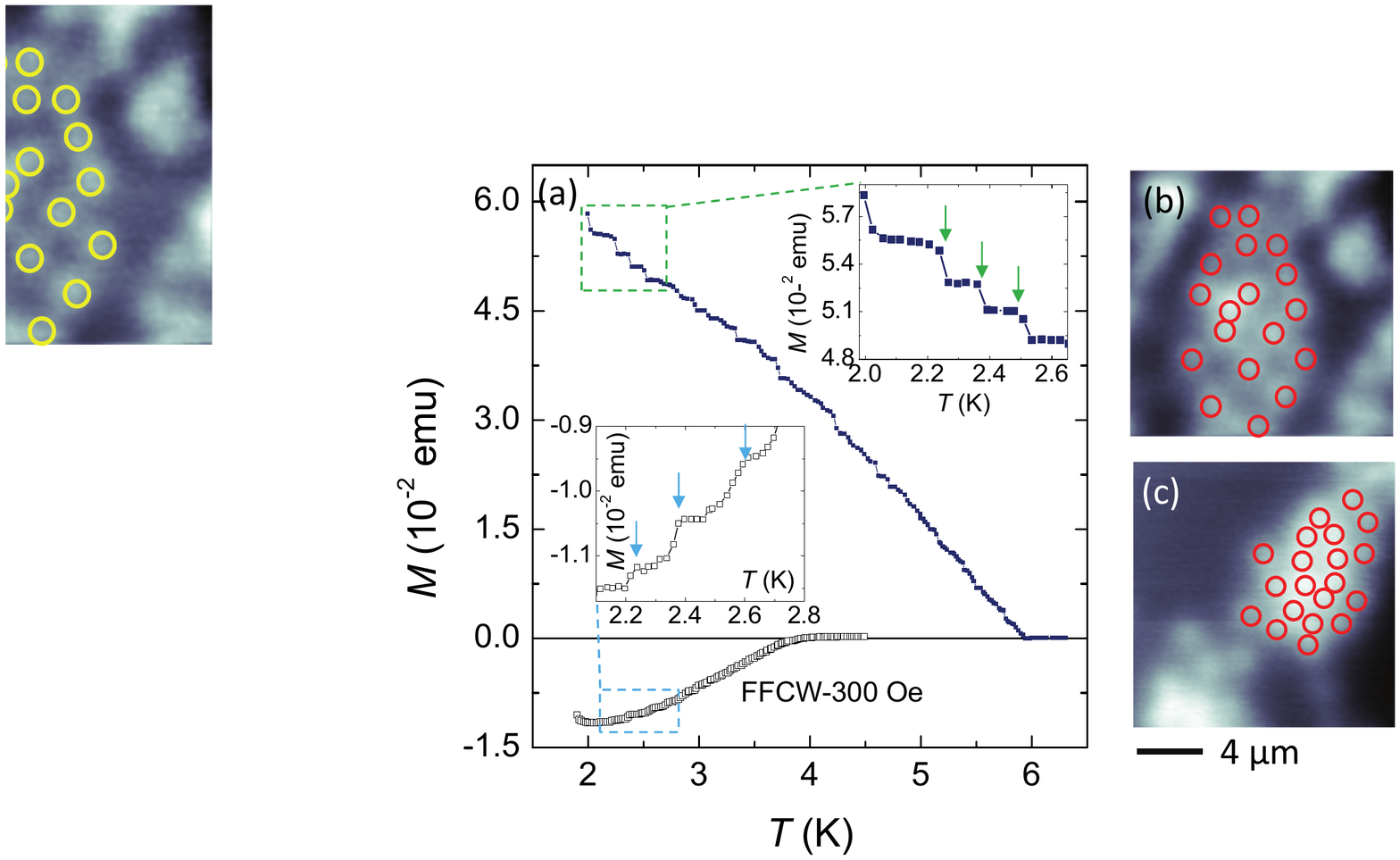}
\caption{(Color online)(a) Temperature dependence of magnetization (filled square) measured after performing rapid cooling at 300 Oe and subsequently removing the magnetic field. The bottom panel shows the magnetization as a function of temperature (open squares) measured in FFCW mode at 300 Oe. The insets show the close-up of the $M(T)$ curves in the low temperature regions as indicated by the dashed rectangles; The arrows in the insets mark the magnetization jumps. (b), (c) SHPM images of typical vortex clusters observed inside the ZrB$_{12}$ sample at 4.2 K after rapid cooling with $H=2$ Oe. The circles indicate the positions of single quantum vortices in the clusters.} \label{fig:2}
\end{figure*}

Fig.1 shows the PME observed in FFCW mode under magnetic fields of 1 Oe, 5 Oe and 10 Oe. Below $T_\textrm{c}$, all the curves show pronounced dips, which are characteristic of the PME at relatively big magnetic fields \cite{Li,Braunisch}. We would like to mention that positive magnetization signal is only observed at relatively small magnetic fields (left inset of Fig.~1), since at higher magnetic fields, the paramagnetic signal from the PME is combined with a relatively strong diamagnetic signal of the superconductor. Thus, the total magnetic moment becomes negative and only an anomalous dip can be observed. With increasing field, the dip moves to lower temperature and considerably broadens. The middle inset of Fig.1 shows the temperature dependence of the magnetization in ZFC and FFCW mode. In ZFC modes, the sample shows the conventional behavior with a critical temperature of 5.95 K. The PME only appears in the FFCW mode. So far, many model have been proposed to explain the PME, including the $\pi$-junctions related to d-wave pairing symmetry in high-$T_c$ superconductors and flux compression mechanism. We have ruled out the former due to the fact that our sample is a traditional BCS superconductor with s-wave paring symmetry \cite{Tsindlekht}. In the following paper, we interpret the data with flux compression model. According to the flux compression mechanism, when performing field-cooling through the third critical temperature (surface superconductivity), the flux is expelled from the surface layer toward the inside region. Due to the confined geometry, this trapped flux forms the vortex clusters with a fixed total vorticity $L$, giving a positive contribution to the magnetization. The right inset of Fig. 1 shows the SFCC curve at $H=50$ Oe. No PME is observed.

To study the behavior of the trapped vortices, we perform a rapid cooling at 300 Oe to 2 K and subsequently remove the field. Due to vortex pinning, the number of the trapped vortices should not change after removing the external field. As seen in the upper panel of Fig.2a, the magnetization jumps to a positive value after the external field is removed. Then the magnetization is measured with increasing temperature. The trapped vortices are expected to be gradually expelled out of the sample interior due to the weakening of the pinning strength at elevated temperatures, where thermal fluctuations become more and more pronounced. From the upper panel (filled squares) it is remarkable to see that the magnetization decreases with step-like jumps especially pronounced at low temperatures (top inset of Fig.2a). Each of the step-like jumps cannot be attributed to the expulsion of a single quantum vortex since the number of the trapped vortices can be up to several thousands and only 40 steps are identified in our case. The step-like jumps are also observed in the $M(T)$ curve when magnetic flux penetrates into the sample under external field (lower inset of Fig. 2a). Due to the relatively weak pinning in our sample \cite{Ge-ZrB12}, the possibility of flux jump in traditional type-II superconductors has also been ruled out. In type-II/1 superconductors, instead of a triangular vortex lattice, vortex clusters form due to the non-monotonic, i.e., long-range attractive and short-range repulsive vortex-vortex interactions. 

The vortex clusters at relatively small magnetic fields are directly visualized by using SHPM. Two typical vortex clusters surrounded by the Meissner phase are shown in Figs. 2b and c. It has been found that the size of vortex clusters increases with cooling field \cite{Ge-ZrB12}. At high magnetic fields, we expect large vortex clusters, which incorporate more vortices. However, due to the limited spacial resolution of the SHPM, we are not be able to directly image them. Further study with scanning tunneling microscopy is needed. In type-II/1 superconductors, each vortex cluster behaves as one object and has repulsive interactions with other vortex clusters\cite{Brandt}. The most plausible scenario to account for the step-like jumps in the $M(T)$ curve is that the penetration and expulsion of magnetic flux are implemented in the form of vortex clusters.  Each step corresponds to the expulsion of a large number of vortex clusters. For type-II/1 superconductors, the non-monotonic vortex-vortex interactions only appear at low temperatures, while at high temperatures close to $T_\textrm{c}$, repulsive vortex-vortex interaction dominates as the sample transits to the traditional type-II/2 regime\cite{Ge-ZrB12,Vagov}. As a result, vortex clusters disappear and a homogeneous distribution of a triangular vortex lattice is formed. This is also consistent with our experiments that the step-like jumps are obviously observed at low temperatures where the samples are well in the type-II/1 regime (bottom panel of Fig. 2a). Close to $T_\textrm{c}$, single quantum vortices play the main role in the expulsion and penetration of magnetic flux.

\begin{figure}[bt]
\centering
\includegraphics*[width=1\linewidth]{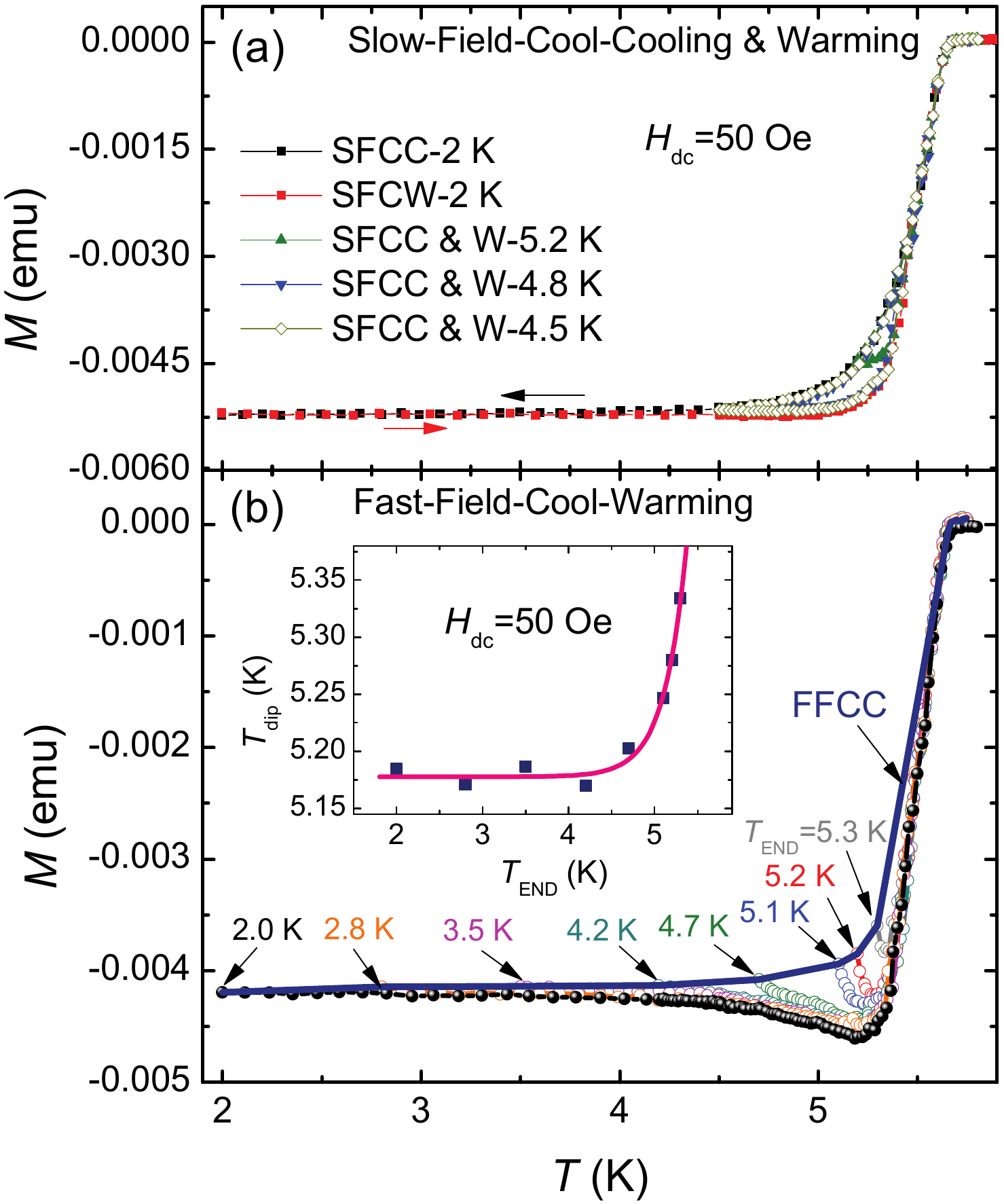}
\caption{(Color online) (a) Temperature dependence of the magnetization measured during field cooling to different desired temperatures and then warming up with a slow rate of 0.03 K/min. (b) Temperature dependence of the magnetization measured after performing rapid cooling (5 K/min) to different temperatures and then warming up. The solid line shows the curve of the FFCC as the envelope of the endpoints of the FFCW data. The inset shows $T_\textrm{dip}$ as a function of $T_\textrm{END}$. The arrows mark the temperature $T_\textrm{END}$ from which the $M(T)$ curves are measured.} \label{fig:3}
\end{figure}

\begin{figure}[t]
\centering
\includegraphics*[width=1\linewidth]{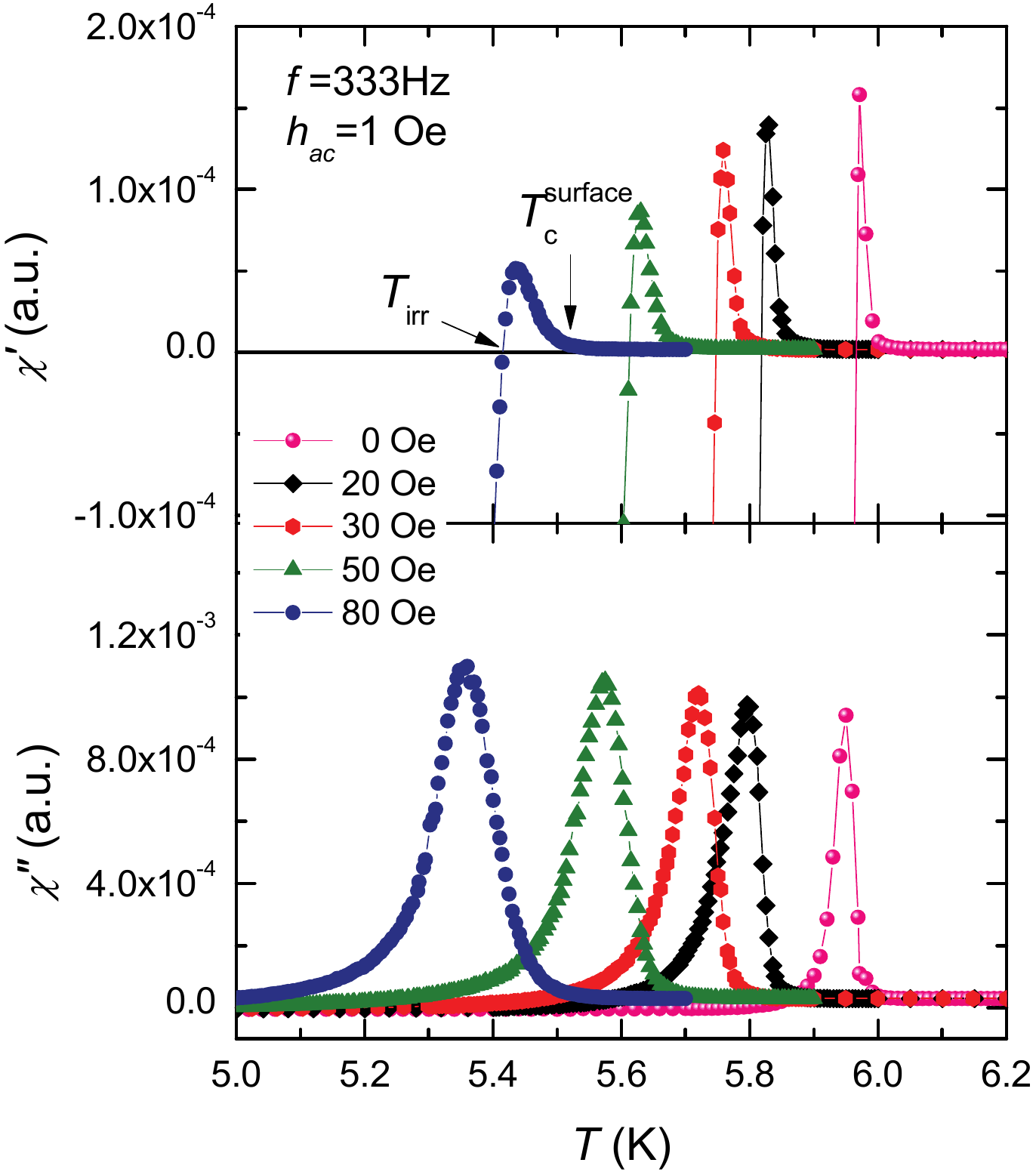}
\caption{(Color online) Temperature dependence of in-phase (upper panel) and out-of-phase (lower panel) ac susceptibility measured at various magnetic fields with an ac field amplitude of  1 Oe and a frequency of 333 Hz. Due to the diffraction paramagnetic effect a peak is observed for each in-phase curve. We identify the onset of DPE as $T^\textrm{surface}_\textrm{c}(H)$, and the onset of a diamagnetic response as $T_\textrm{irr}(H)$.} \label{fig:4}
\end{figure}

\begin{figure*}[htb]
\centering
\includegraphics*[width=0.6\linewidth]{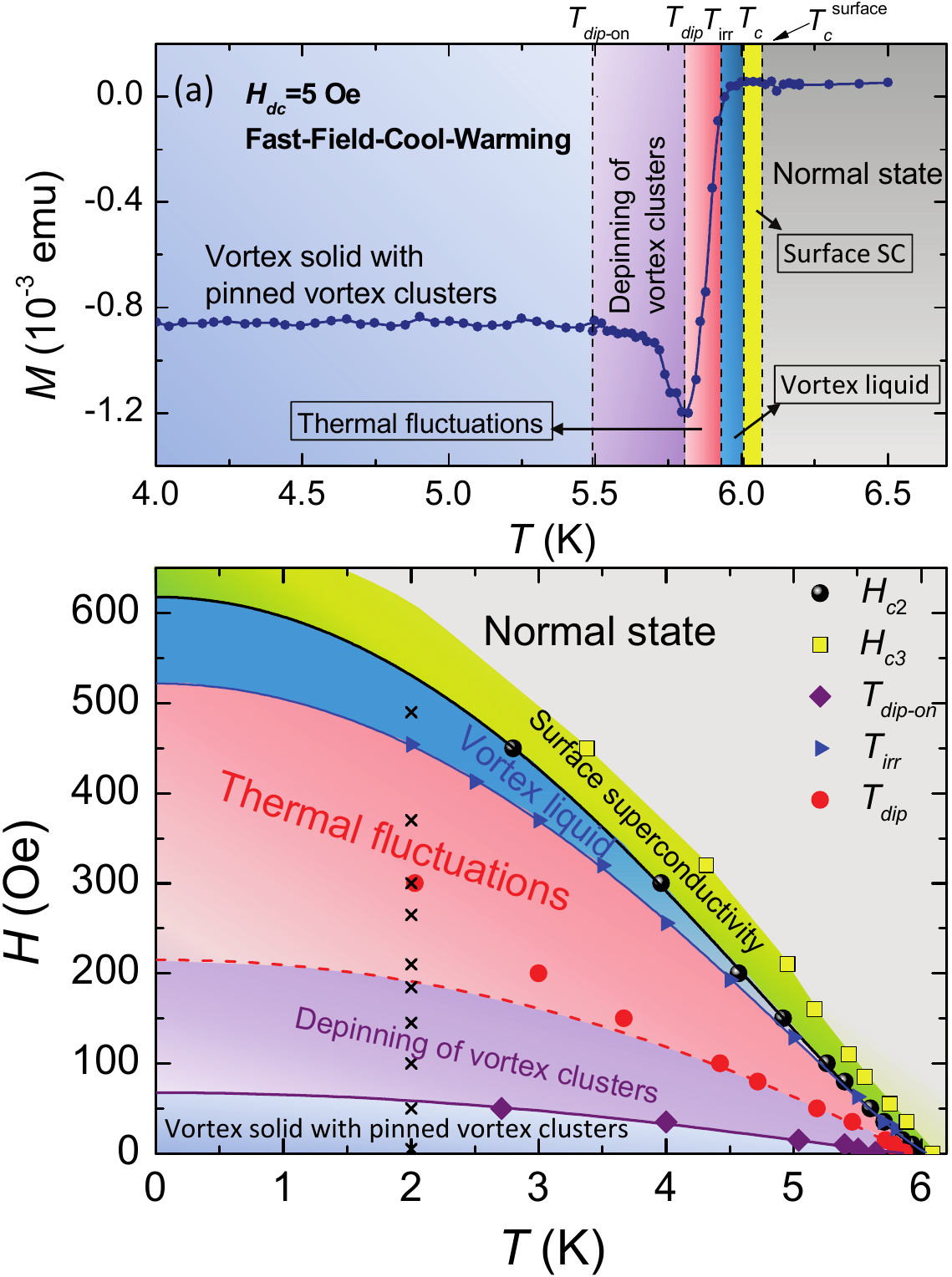}
\caption{(Color online) (a) Temperature dependence of the magnetization in the FFCW mode at 5 Oe. (b) Phase diagram for the PME in FFCW mode. The third critical field $H_\textrm{c3}$ is defined from the onset of DPE in the in-phase ac susceptibility measurements (see Fig. 4); $H_\textrm{c2}$ is defined from the intersection of two linear fits of the $M(T)$ curves above and below the onset; $T_\textrm{irr}$ is derived as the onset of a diamagnetic signal on the in-phase ac susceptibility curve. The solid and dashed lines are plots of the empirical formula $H(T)=H(0)[1-(T/T_\textrm{c})^2]^n$. The crosses show the field locations where the magnetic relaxation curves are measured.} \label{fig:4}
\end{figure*}

In a system with complex interactions (vortex-pinning, vortex-vortex, thermal fluctuations), memory effects often appear \cite{Ge-peak,Pasquini,Paltiel,Valenzuela}. 
We have found that the PME only appears when cooling down the sample at sufficiently high cooling rate (5 K/min). At small cooling rate (0.03 K/min), the extra flux trapped through surface superconductivity has enough time to escape from the sample interior due to flux diffusion, resulting in a stable and more ordered vortex state at low temperatures. Thus when increasing the temperature, there will be no extra magnetic flux expelled from the sample and the anomalous dip below $T_\textrm{c}$ disappears. In Fig. 3a we show the magnetization measured with both decreasing and increasing temperature at 0.03 K/min. The $M(T)$ curves measured during sweeping up the temperature overlap and no magnetization dip is observed. The hysteresis of SFCC and SFCW curves indicates that with decreasing temperature the trapped magnetic flux is gradually expelled from the sample. 

Now we focus on the vortex behavior at fast cooling rate. Since the anomalous dip is related to the metastability of trapped flux, the temperature, $T_\textrm{dip}$, where the dip appears in the FFCW curves can be affected by many parameters. We have found that $T_\textrm{dip}$ depends on the temperature $T_\textrm{END}$ where we start to warm up the sample. In Fig. 3b we perform rapid cooling to different temperatures, and subsequently measure the magnetization with increasing temperature. All the FFCW curves show obvious dips. $T_\textrm{dip}$ is found to follow an exponential change with $T_\textrm{END}$ as shown in the inset of Fig. 3b. Below 4.2 K, $T_\textrm{dip}$ is constant ($\approx$5.18 K). This is understandable since at low temperatures, the pinning force becomes dominant compared with the thermal fluctuation and the elastic force between vortices. The sample is in the solid phase and all the vortices are well pinned. Thus the initial vortex state is the same for all the $M(T)$ curves with low $T_\textrm{END}$. With increasing temperature, the trapped magnetic flux escapes from the sample when some of the vortices are depinned at $T_\textrm{dip-on}$ and finally  a pronounced magnetization dip forms around the temperature $T_\textrm{dip}$.  This results in a similar $T_\textrm{dip}$ for all the curves with low $T_\textrm{END}$. However, after rapid cooling to a higher $T_\textrm{END}$, the trapped extra flux does not have enough time to escape from the sample to reach a stable state. In this case, vortices keep on escaping from the sample.  $T_\textrm{dip}$ is observed to move to higher temperatures. So in order to correctly compare the PME at various magnetic fields, the initial vortex state must be in the vortex solid phase, i.e. all the vortices have to be pinned. 

\begin{figure*}[htb]
\centering
\includegraphics*[width=0.75\linewidth]{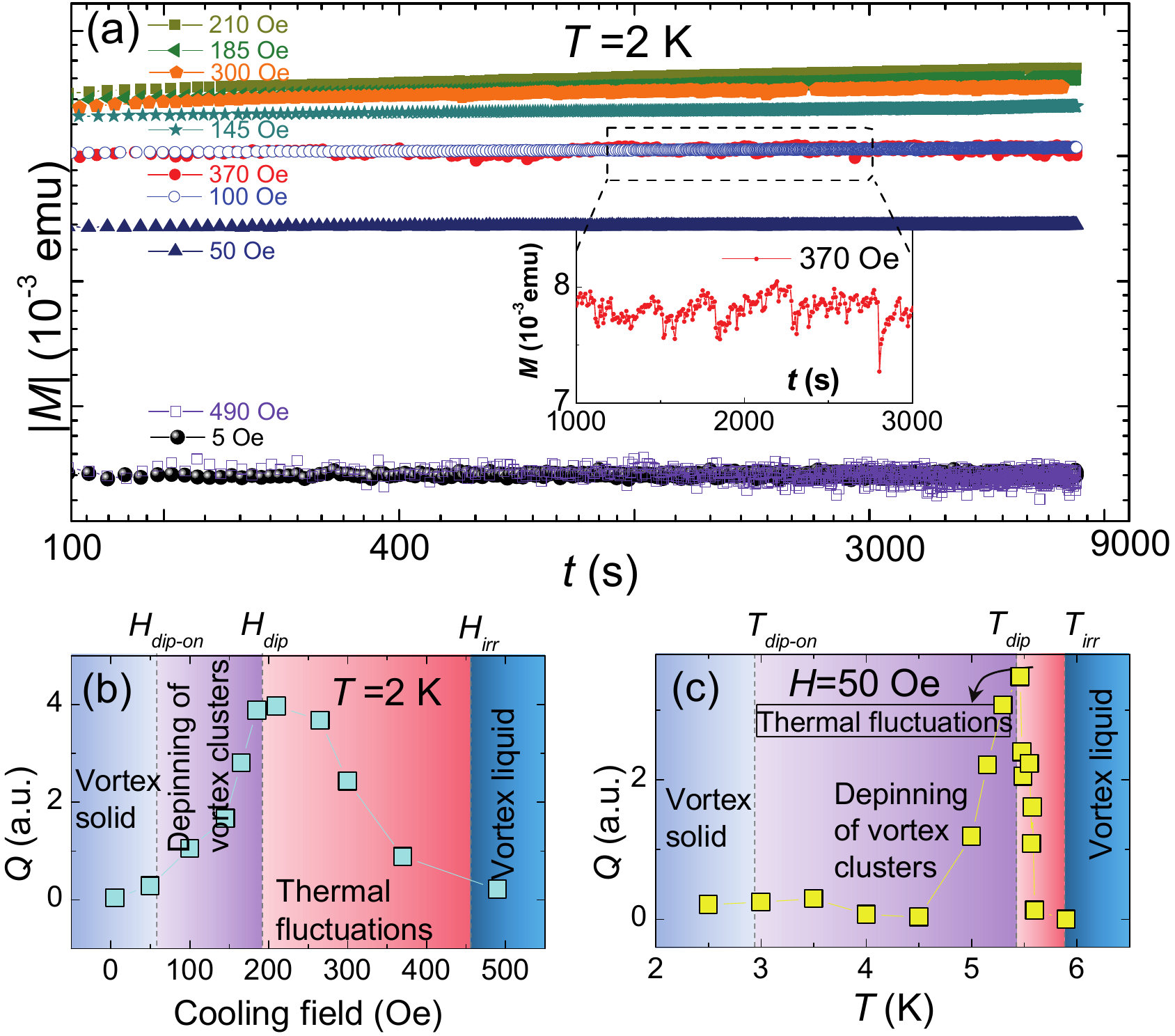}
\caption{(Color online) (a) Magnetic relaxation after rapid cooling to 2 K under various magnetic fields. The inset shows the close-up of magnetic relaxation at 370 Oe. Strong oscillations are observed. (b) Relaxation rate parameter as a function of cooling field at 2 K. (c) Relaxation rate parameter as a function of temperature under the field of 50 Oe. In both (b) and (c), the phase boundaries (dashed lines) are determined from the phase diagram of Fig. 5b.} \label{fig:5}
\end{figure*}

Based on the analysis above, the PME is determined by the interplay between three energies: the pinning energy; the vortex (cluster) elastic energy and the thermal fluctuation energy. After rapid cooling, the trapped flux in the form of vortex clusters is well pinned inside the sample, resulting in a rather inhomogeneous distribution of vortices so that part of the sample remains in the Meissner state while the rest is penetrated by high density magnetic flux (see Fig. 2b and c). When increasing temperature up to $T_\textrm{dip-on}$, the pinning force becomes comparable to the repulsion between vortex clusters. In order to form a configuration that corresponds to a thermodynamically stable state, the mobile vortices (vortex clusters) tend to reduce the ``excess flux'' (``excess'' as compared to that in the thermodynamic equilibrium). Then the vortex clusters, corresponding to the``excess'' trapped flux, start to be expelled out of the sample interior as already discussed in relation to Fig. 2. Thus, the magnetization signal becomes more negative and a dip appears at $T_\textrm{dip}$. When the temperature is increased further above $T_\textrm{dip}$, the thermal fluctuations become more and more evident and external magnetic flux can also penetrate into the sample. 
Just below $T_\textrm{c}$ the sample transits to the vortex liquid state, where all the vortices in the sample can move freely. In this region, the penetration and expulsion of flux will be in the form of single quantum vortices.

In order to determine the irreversibility and surface superconductivity phase boundary, we measured the temperature dependence of the ac susceptibility of the sample. As shown in Fig. 4, the existence of a reversible magnetization response close to the transition temperature yields a paramagnetic signal on the in-phase ac susceptibility curve. The anomalous peak, termed as diffraction paramagnetic effect (DPE), has been used to determine precisely the irreversibility line \cite{DPE}. Here, the surface superconductivity transition temperature $T^\textrm{surface}_\textrm{c}(H)$ and the irreversibility temperature $T_\textrm{irr}(H)$ are determined as the onset of DPE and the onset of diamagnetic response, respectively. 

In Fig. 5a we show one typical magnetization vs. temperature curve in the FFCW mode at 5 Oe. The critical parameters discussed above are indicated. The whole temperature region is divided into a few phases with different vortex configurations. By repeating the measurements at various external fields, we are able to construct the $H-T$ phase diagram for the DPE. In Fig. 5b, the data are displayed by the symbols, while the solid lines are fit with the empirical formula $H(T)=H(0)[1-(T/T_\textrm{c})^2]^n$ with $n$ a fitting parameter. We observe that, for $H_\textrm{c2}$, $T_\textrm{irr}$, $T_\textrm{dip}$ and $T_\textrm{dip-on}$, $n$ equals 1.34, 1.26, 1.46 and 1.14, respectively. 
It should be noted that the vortex solid phase only occupies a small region in the phase diagram. Above $H=67$ Oe, it is impossible for the pinning centers to pin all the trapped flux. Hence, the measured PME is not reliable and  $T_\textrm{dip}$ deviates from the fitting curve (dashed line) which is based on the data below $H=50$ Oe. Also compared with vortex phase diagram constructed through flux penetration process \cite{Ge-ZrB12}, no Meissner state is recovered even at magnetic fields much smaller than $H_{c1}$ \cite{Ge-NJP}. This is mainly due to the pinning and surface barrier of the sample. We suggest that our phase diagram is applicable for but not restricted to type-II/1 superconductors. For type-II/2 superconductors, a similar phase diagram is expected. The only difference is that, in type-II/2 superconductors, the expulsion and penetration of magnetic flux will happen through individual vortices instead of vortex clusters. 

It has been predicted that the non-monotonic vortex interactions may facilitate the trapping of magnetic flux, thus lead to giant PME \cite{Silva}. However, our experimental results revealed that no giant PME is observed in ZrB$_{12}$ with non-monotonic vortex-vortex interaction. This may suggest that further corrections may be needed for the theories used in Ref. \cite{Silva}. In fact, the non-monotonic vortex interactions only appears in type-II/1 superconductors at low temperatures, while at high temperatures close to $T_c$, the vortex interaction is purely repulsive (in type-II/2 regime). So when cooling down the type-II/1 superconductor, the sample has to first go through the type-II/2 regime. For type-II/1 and type-II/2 superconductors, if the pinning strength, sample geometries, thermal fluctuations are the same, then when quench the samples, the trapped amount of magnetic flux should be similar, since the initial vortex-vortex interactions close to $T_c$ are both repulsive. 

One efficient way to study a system with metastable magnetic responses is to measure the magnetic relaxation with time. This has been used to study spin dynamics \cite{Lundgren,Paulsen}, interacting magnetic nanoparticles \cite{Jonsson,Mamiya} and superconductors \cite{Papadopoulou,Assi,Wen}. In order to probe the vortex configurations in different regions of the phase diagram, we measured the magnetic relaxation across the different phase boundaries as indicated by the crosses in Fig. 5b. It should be noted that, for a superconductor cooled down with a low cooling rate, the magnetization reaches its equilibrium and no relaxation can be detected. However, in superconductors with PME, an extra magnetic flux are trapped inside the superconductor after fast-field-cooling. The relaxation measurements will provide valuable information about the dynamics of the extra flux. In Fig. 6a we show the magnetic relaxation results obtained by performing rapid cooling at various fields to 2 K and then recording the magnetization with an observation time up to 2 hours. On a long-time scale,  $t>t_0$ with $t_0$ of the order of $10^2$ s, the logarithm of the magnetization shows a linear dependence on the logarithm of the observation time. We define the magnetic relaxation-rate parameter $Q=\partial(lnM)/\partial(lnt)$. As shown in Fig. 6b, at low and high enough fields the relaxation-rate parameter $Q$ is quite small. Only in the middle range of fields, the relaxation is clearly observed. This is consistent with the phase diagram discussed above. At low fields, the sample is in the vortex solid state with all the vortices being well pinned. Thus no magnetic relaxation occurs. With increasing field, the density of vortices and their mutual repulsion increase, the pinning forces become less dominant and part of the vortices can escape from the sample interior. A peak in the relaxation rate parameter appears around 200 Oe. At higher fields, i.e., in the fluctuation-dominated regime, the role of pinning is strongly reduced and an increasing number of vortices (clusters) are highly mobile, the relaxation to the metastable state occurs on a very short time scale. As a result, the relaxation rate decreases. Due to strong thermal fluctuations vortices are able to penetrate into the sample. This is also evidenced by the strong fluctuations of the magnetization in this region as shown in the inset of Fig.6a.  Above the irreversibility line the sample enters the vortex liquid phase. The penetration and expulsion of vortices can reach equilibrium quickly. Thus, the relaxation rate parameter goes to zero. We also measured (see Fig. 5c) the relaxation rate as a function of temperature at 50 Oe horizontally across the phase diagram. A similar correlation between the behavior of $Q$ and the phase encountered is observed.

\section{IV. CONCLUSION}
In summary, we have studied the paramagnetic Meissner effect in a ZrB$_{12}$ single crystal with $\kappa \approx 0.8 $ which is between the traditional type-I and type-II regime. PME characterized by a negative dip of the magnetization is observed in a ZrB$_{12}$ single crystal after rapid cooling. It is found that the expulsion and penetration of flux are in the form of vortex clusters at low temperatures. We proposed that the observed PME can be interpreted in terms of the interplay  among the flux compression, the different temperature dependencies of the vortex-vortex and the vortex-pin interactions, and thermal fluctuations. A detailed $H$-$T$ phase diagram is constructed for the PME. The relaxation rate as a function of cooling field and temperature correlate well with the phase diagram. 

\section{ACKNOWLEDGMENTS}
We acknowledge support from the Methusalem funding of the Flemish government, the Flemish Science Foundation (FWO-Flanders) and the MP1201 COST action.


\begin{references}

\bibitem{Li} M. S. Li, Phys. Rep. {\bf 376}, 133 (2003).
\bibitem{Svedlindh} P. Svedlindh, K. Niskanen, P. Norling, P. Nordblad, L. Lundgren, B. Lönnberg, and T. Lundstrom, Physica C {\bf 1365}, 162 (1989).
\bibitem{Sigrist} M. Sigrist and T. M. Rice, J. Phys. Soc. Jpn. {\bf 61}, 4283 (1992); Rev. Mod. Phys. {\bf 67}, 503 (1995).
\bibitem{Thompson} D. J. Thompson,  M. S. M. Minhaj, L. E. Wenger, and J. T. Chen,  Phys. Rev. Lett. {\bf 75}, 529 (1995).
\bibitem{Kostic} P. Kostic, B. Veal, A. P. Paulikas, U. Welp, V. R. Todt, C. Gu, U. Geiser, J. M. Williams, K. D. Carlson, and R. A. Klemm, Phy. Rev. B {\bf 53}, 791 (1996).
\bibitem{Geim} A. K. Geim, S. V. Dubonos, J. G. S. Lok, M. Henini, and J. C. Maan, Nature {\bf 396}, 144 (1998).
\bibitem{Koshelev} A. E. Koshelev and A. I. Larkin, Phys. Rev. B {\bf 52}, 13559 (1995).
\bibitem{Moshchalkov} V.V. Moshchalkov, X.G. Qiu, and V. Bruyndoncx, Phys. Rev. B {\bf 55}, 11793 (1997).
\bibitem{Okram} G. S. Okram, D. T. Adroja, B. D. Padalia, O. Prakash, and P. A. J. de Groot, J. Phys.: Condens. Matter {\bf 9}, L525 (1997).
\bibitem{Felner} I. Felner, M. I. Tsindlekht, G. Drachuck, A. Keren, J. Phys.: Condens. Matter {\bf 25}, 065702 (2013).
\bibitem{Silva} R. M. da Silva, M. V. Milosevic, A. A. Shanenko, F. M. Peeters and J. Albino Aguiar, Sci. Rep. {\bf 5}, 12695 (2015).
\bibitem{Sluchanko} N. E. Sluchanko, A. N. Azarevich, A. V. Bogach, S. Yu. Gavrilkin, V. V. Glushkov, S. V. Demishev, A. V. Dukhnenko, A. B. Lyashchenko, K. V. Mitsen, and V. B. Filipov, JETP Lett. {\bf 94}, 642-646 (2011).
\bibitem{Ge-ZrB12} J.-Y. Ge, J. Gutierrez, A. Lyashchenko, V. Filipov, J. Li, and Victor V. Moshchalkov, Phys. Rev. B {\bf 90}, 184511 (2014).
\bibitem{Leviev} G. I. Leviev, V. M. Genkin, M. I. Tsindlekht, I. Felner, Y. B. Paderno, and V. B. Filippov, Phys. Rev. B {\bf 71}, 064506 (2005).
\bibitem{Tsindlekht} M. I. Tsindlekht,  G. I. Leviev, V. M. Genkin, I. Felner, Y. B. Paderno, and V. B. Filippov, Phys. Rev. B {\bf 73}, 104507 (2006).
\bibitem{Wang} Y. Wang, R. Lortz, Y. Paderno, V. Filippov, S. Abe, U. Tutsch, and A.Junod, Phys. Rev. B {\bf 72}, 024548 (2005).
\bibitem{Daghero} D. Daghero, R. S. Gonnelli, G. A. Ummarino, A. Calzolari, V. Dellarocca, V. A. Stepanov, V. B. Filippov, and Y. B. Paderno, Supercond. Sci. Technol. {\bf 17}, S250 (2004).
\bibitem{Braunisch} W. Braunisch, N. Knauf, G. Bauer, A. Kock, A. Becker, B. Freitag, A. Grütz, V. Kataev, S. Neuhausen, B. Roden, D. Khomskii, D. Wohlleben, J. Bock, and E. Preisler, Phys. Rev. B {\bf 48}, 4030 (1993).
\bibitem{Tsindlekht} M. I. Tsindlekht, G. I. Leviev, I. Asulin, A. Sharoni, O. Millo, I. Felner, Yu. B. Paderno, V. B. Filippov, and M. A. Belogolovskii, Phys. Rev. B {\bf 69}, 212508 (1993).
\bibitem{Ge-NC} J.-Y. Ge, J. Gutierrez, V. N. Gladilin,  J. T. Devreese, and V. V. Moshchalkov, Nat. Commun. {\bf 6}, 6573 (2015).
\bibitem{Ge-NC2} J.-Y. Ge, V. N. Gladilin, J. Tempere, C. Xue, J. T. Devreese, J. Van de Vondel, Y. Zhou, and V. V. Moshchalkov, Nat. Commun. {\bf 7}, 13880 (2016).
\bibitem{Brandt} E. H. Brandt and M. P. Das, J. Supercond. Nov. Magn. {\bf 24}, 57 (2011).
\bibitem{Vagov} A. Vagov, A. A. Shanenko, M. V. Milosevic, V. M. Axt, V. M. Vinokur, J. A. Aguiar, and F. M. Peeters, Phys. Rev. B {\bf 93}, 174503 (2016).
\bibitem{Ge-peak} J.-Y. Ge, J. Gutierrez, J. Li, J. Yuan, H.-B. Wang, K. Yamaura, E. Takayama-Muromachi, and V. V. Moshchalkov, Phys. Rev. B {\bf 88}, 144505 (2013).
\bibitem{Pasquini} G. Pasquini, D. P. Daroca, C. Chiliotte, G. S. Lozano, and V. Bekeris, Phys. Rev. Lett. {\bf 100}, 247003 (2008).
\bibitem{Paltiel} Y. Paltiel, E. Zeldov, Y. N. Myasoedov, H. Shtrikman, S. Bhattacharya, M. J. Higgins, Z. L. Xiao, E. Y. Andrei, P. L. Gammel, and D. J. Bishop, Nature (London) {\bf 403}, 398 (2000).
\bibitem{Valenzuela} S. O. Valenzuela and V. Bekeris, Phys. Rev. Lett. {\bf 84}, 4200 (2000).
\bibitem{DPE} S. Ramakrishnan, R. Kumar, P. L. Paulose, A. K. Grover, and P. Chaddah, Phys. Rev. B {\bf 44}, 9514 (1991).
\bibitem{Ge-NJP} J. Ge, J. Gutierrez, B. Raes, J. Cuppens, and V. V. Moshchalkov, New J. Phys. {\bf 15}, 033013 (2013).
\bibitem{Lundgren} L. Lundgren, P. Svedlindh, P. Nordblad, and O. Beckman, Phys. Rev. Lett. {\bf 51}, 911 (1983).
\bibitem{Paulsen} C. Paulsen,	M. J. Jackson,	E. Lhotel,	B. Canals,	D. Prabhakaran, K. Matsuhira, S. R. Giblin, and S. T. Bramwell, Nature Phys. {\bf 10}, 135-139 (2014).
\bibitem{Jonsson} T. Jonsson, J. Mattsson, C. Djurberg, F. A. Khan, P. Nordblad, and P. Svedlindh, Phys. Rev. Lett. {\bf 75}, 4138 (1995).
\bibitem{Mamiya} H. Mamiya, I. Nakatani, and T. Furubayashi, Phys. Rev. Lett. {\bf 88}, 067202 (2002).
\bibitem{Papadopoulou} E. L. Papadopoulou, P. Nordblad, P. Svedlindh, R. Schoneberger, and R. Gross, Phys. Rev. Lett. {\bf 82}, 173 (1999).
\bibitem{Assi} H. Assi, H. Chaturvedi, U. Dobramysl, M. Pleimling, and U. C. Tauber, Phys. Rev. E {\bf 92}, 052124 (2015).
\bibitem{Wen} H. Yang, C. Ren, L. Shan, and H.-H. Wen, Phys. Rev. B {\bf 78}, 092504 (2008).

\end{references}
\end{document}